\documentclass{aa}  

\usepackage{graphicx}
\usepackage{txfonts}
\usepackage{lipsum}
\usepackage{subcaption}
\usepackage{lscape}
\usepackage{placeins}

\begin{document}

\title{Unsupervised machine learning classification of gamma-ray bursts based on the rest-frame prompt emission parameters}

\titlerunning{Unsupervised Machine Learning Classification of Gamma-Ray Bursts}

\author{Si-Yuan Zhu\inst{1,2}
	\and Lang Shao\inst{3}
    \and Pak-Hin Thomas Tam\inst{2}
    \and Fu-Wen Zhang\inst{1,4}\fnmsep\thanks{Corresponding authors: fwzhang@pmo.ac.cn}
    }

\institute{College of Physics and Electronic Information Engineering, Guilin University of Technology, Guilin 541004, China\\
         \and School of Physics and Astronomy, Sun Yat-Sen University, Zhuhai 519082, China\\
         \and Department of Space Sciences and Astronomy, Hebei Normal University, Shijiazhuang 050024, China\\
         \and Key Laboratory of Low-dimensional Structural Physics and Application, Education Department of Guangxi Zhuang Autonomous Region, Guilin 541004, China}

\date{Received September 30, 20XX}

\abstract
{Gamma-ray bursts (GRBs) are generally believed to originate from two distinct progenitors, compact binary mergers and massive collapsars.
Traditional and some recent machine learning-based classification schemes predominantly rely on observer-frame physical parameters, which are significantly affected by the redshift effects and may not accurately represent the intrinsic properties of GRBs.
In particular, the progenitors usually could only be decided by successful detection of the multi-band long-term afterglow, which could easily cost days of devoted effort from multiple global observational utilities.
In this work, we apply the unsupervised machine learning (ML) algorithms called t-SNE and UMAP to perform GRB classification based on rest-frame prompt emission parameters.
The map results of both t-SNE and UMAP reveal a clear division of these GRBs into two clusters, denoted as GRBs-I and GRBs-II.
We find that all supernova-associated GRBs, including the atypical short-duration burst GRB 200826A (now recognized as collapsar-origin), consistently fall within the GRBs-II category.
Conversely, all kilonova-associated GRBs (except for two controversial events) are classified as GRBs-I, including the peculiar long-duration burst GRB 060614 originating from a merger event.
In another words, this clear ML separation of two types of GRBs based only on prompt properties could correctly predict the results of progenitors without follow-up afterglow properties.
Comparative analysis with conventional classification methods using $T_{90}$ and $E_{\rm p,z}$--$E_{\rm iso}$ correlation demonstrates that our machine learning approach provides superior discriminative power, particularly in resolving ambiguous cases of hybrid GRBs.}


\maketitle

\section{Introduction} \label{sec:introduction}
Gamma-ray bursts (GRBs) are the brightest electromagnetic explosions in the universe.
Theoretically, GRBs may have two types of origins, one originating from compact binary merger (Type I GRBs) and the other from the collapse of massive stars \citep[Type II GRBs;][]{1986ApJ...308L..43P,1993ApJ...417L..17K,1993ApJ...405..273W,1999A&A...344..573R,2001ApJ...550..372F}.
The durations of GRBs originating from collapsars are determined by the envelope fallback timescale, which typically spans around 10 seconds.
While the timescales of GRBs originating from NS--NS or NS--BH mergers are typically 0.01--0.1 s, suggesting that Type I GRBs are generally shorter in duration than Type II GRBs \citep{2005A&A...436..273A}.

According to the bimodal distribution of the GRB durations observed by BATSE, GRBs are phenomenally classified as long GRBs (LGRBs) with duration longer than 2 seconds ($T_{90}>2$ s) and short GRBs (SGRBs) with duration shorter than 2 seconds \citep[$T_{90}<2$ s;][]{1993ApJ...413L.101K}.
Observationally, some LGRBs were associated with Type Ic supernovae (SNe), such as GRB 980425/SN 1998bw and GRB 030329/SN 2003dh \citep{1998Natur.395..670G,2003ApJ...591L..17S}.
It is believed that LGRBs may originate from the core-collapse of massive stars.
Moreover, GRB 170817A, a SGRB observed by Fermi and INTEGRAL, is associated with gravitational wave (GW) GW170817 and kilonova (KN) AT 2017gfo, which has confirmed that a part of SGRBs originate from the merging of binary compact stars \citep{2017PhRvL.119p1101A,2017ApJ...848L..14G,2017ApJ...848L..15S,2017ApJ...851L..18W}.

However, recent observations suggest that a short duration GRB 200826A was observed to associate with a SN \citep{2021NatAs...5..917A,2021NatAs...5..911Z,2022ApJ...932....1R}, as well as long duration GRBs 060614, GRB 211211A, and GRB 230307A are associated with KNe \citep{2015ApJ...811L..22J,2015NatCo...6.7323Y,2022Natur.612..232Y,2022Natur.612..223R,2024Natur.626..737L,2024Natur.626..742Y}.
These peculiar GRBs have severely broken the associations between SGRBs and Type I GRBs, as well as between LGRBs and Type II GRBs, respectively.
Therefore, the classification of GRBs based on duration alone is unreliable.

The $E_{\rm p,z}$--$E_{\rm iso}$ correlation has been widely discussed as an another GRB classification indicator due to the
fact that SGRBs/Type I GRBs and LGRBs/Type II GRBs generally follow different $E_{\rm p,z}$--$E_{\rm iso}$ correlations, where $E_{\rm p,z}$ is the rest-frame peak energy in $\nu f_\nu$ spectrum and $E_{\rm iso}$ is the isotropic energy of the prompt emission of GRBs \citep{2002A&A...390...81A,2010ApJ...725.1965L,2012ApJ...750...88Z,2013MNRAS.430..163Q,2020MNRAS.492.1919M,2023MNRAS.524.1096L,2023ApJ...950...30Z,2024ApJ...976...62Z}.
It is worth noting that the low-luminosity LGRBs (such as GRB 980425 and GRB 131203) are the obvious outliers of the $E_{\rm p,z}$--$E_{\rm iso}$ correlation followed by the typical LGRBs.
In addition, $E_{\rm p,z}$--$E_{\rm iso}$ correlations for SGRBs and LGRBs become overlapped as the sample of GRBs increases.
Using the $E_{\rm p,z}$--$E_{\rm iso}$ correlations, \cite{2010ApJ...725.1965L} and \cite{2020MNRAS.492.1919M} also proposed two similar classification methods by introducing two parameters, defined as $\varepsilon$ and $EH$, $\varepsilon = E_{\rm iso,52}\ E_{\rm p,z,2}^{-5/3}$ and $EH = E_{\rm p,z,2}\ E_{\rm iso,51}^{-0.4}$, respectively.
Considering the duration, \cite{2020MNRAS.492.1919M} proposed the energy-hardness-duration ($EHD$) parameter, $EHD = E_{\rm p,z,2}\ E_{\rm iso,51}^{-0.4}\ T_{\rm 90,z}^{-0.5}$, to classify GRBs.
However, these classification methods cannot identify several special GRBs (such as long Type I GRB 060614, GRB 211211A and GRB 230307A).

Moreover, it is found that magnetar giant flares (MGFs), if occurring in nearby galaxies, would appear as cosmic short-hard GRBs \citep{2020ApJ...899..106Y,2021ApJ...907L..28B,2021Natur.589..207R,2021Natur.589..211S,2024Natur.629...58M,2024A&A...687A.173T,2024ApJ...963L..10Y}.
The light curves and spectra of MGFs are not significantly different from typical SGRBs.
Compelling evidence for period-modulated tail induced by the magnetar rotation cannot be observed by current wide-field gamma-ray monitors \citep{2021Natur.589..211S}.
Thus, the main evidence for distinguishing MGFs from SGRBs is the spatial coincidence with nearby galaxies and the consistency of the energetics with known MGFs.
MGFs may also follow the $E_{\rm p,z}$--$E_{\rm iso}$ correlation with the same slope as LGRBs and SGRBs but with significantly different intercepts \citep{2020ApJ...899..106Y,2020ApJ...903L..32Z,2024ApJ...963L..10Y}.
This makes the classification of GRBs even more confusing.

Recently, machine learning has been widely applied to classify GRBs, containing unsupervised and supervised algorithms.
Especially, the t-distributed Stochastic Neighbor Embedding (t-SNE) and the Uniform Manifold Approximation and Projection (UMAP) are the two most popular unsupervised machine learning algorithms \citep{2008JMLR.9.2579M,2014JMLR.15.3221M,2018arXiv180203426M}.
Using t-SNE or/and UMAP, some studies classified GRBs into two or five types based on GRB light curves and spectra \citep{2020ApJ...896L..20J,2023ApJ...945...67S,2023ApJ...949L..22D,2024MNRAS.527.4272C,2024ApJ...974...55D,2025ApJ...981...14N}.
Using the four parameters (duration, peak energy, fluence, and peak flux) of GRB prompt emission, \cite{2024MNRAS.532.1434Z} also obviously classified GRBs into two clusters.
Especially, a peculiar short GRB 200826A associated with a SN can be well classified into the cluster of typical Type II.
The potential drawback with the previous methods is that they are all based on physical parameters in the observer frame, which is severely affected by the redshift effect.
The physical parameters in the rest frame can reflect the intrinsic properties of GRBs.

In this work, we collect all the GRBs with known redshift and well-measured $E_{\rm p}$ and apply t-SNE and UMAP methods to classify these GRBs based on the prompt emission parameters in the rest frame.
The structure of this paper is organized as follows.
In Section \ref{sec:data}, we describe t-SNE and UMAP methods, as well as the sample selection, respectively.
The classification results of the rest frame sample are shown in Section \ref{sec:rest}.
The discussion for the classification results are shown in Section \ref{sec:discussions}.
The conclusions are shown in Section \ref{sec:conclusions}.
Throughout this paper, the symbolic notation $Q_{\rm n} = Q/10^{\rm n}$ is adopted.

\section{Method and data} \label{sec:data}
\subsection{Method} \label{subsec:method}
t-SNE and UMAP are two unsupervised machine learning algorithms that can nonlinearly reduce high-dimensional data to two-dimensional or three-dimensional for visualization \citep{2008JMLR.9.2579M,2014JMLR.15.3221M,2018arXiv180203426M}.
The basic principles of both t-SNE and UMAP are to embed similar data points in higher dimensional space into similar positions in lower dimensional space, while dissimilar data points are embedded in separate positions in lower dimensional space.
The coordinates resulting from t-SNE and UMAP dimensionality reduction do not have proper labels or physical meaning, but can be very conveniently shown on x-axis and y-axis, respectively, as in a two-dimensional case.
For more details on the t-SNE and UMAP methods please refer to \cite{2018arXiv180203426M} and the references therein.
We adopt both methods so their results can be closely compared for a consistency check.

The python module $scikit$-$learn$\footnote{https://scikit-learn.org/} is applied to the t-SNE analysis.
The most important hyperparameter of t-SNE is $perplexity$, which determines the sizes of the neighborhoods based on the density of the data in the respective regions.
A higher $perplexity$ value will consider having the point have a larger number of neighbors, emphasizing the global structure of the data, while a lower $perplexity$ value will better reflect the local structure of the data.
Choosing the appropriate $perplexity$ value is crucial for representing the different data structures.

The python module $umap$-$learn$\footnote{https://pypi.org/project/umap-learn/} introduced by \citep{2018arXiv180203426M} is applied to the UMAP analysis.
The most important hyperparameters of UMAP are $n\_neighbors$ and $min\_dist$.
$n\_neighbors$ determine the number of neighboring points used in local approximations of manifold structure, and $min\_dist$ controls how tight the embedding is allowed to compress points together.
Note that when t-SNE and UMAP are applied to the dataset, the dimensions of all features must also be the same.
The dataset is standardized using the Z-score standardization method prior to applying t-SNE and UMAP.

\subsection{Data} \label{subsec:data}
To explore the GRB classification in the rest frame, a GRB sample with observed redshift and well-measured spectrum is needed.
\cite{2020MNRAS.492.1919M} compiled an extensive GRB catalog (M20) with redshift, containing $T_{\rm 90,z}$, $E_{\rm p,z}$, and $E_{\rm iso}$.
The major part of our GRB sample has 300 GRBs collected from \cite{2020MNRAS.492.1919M} and its erratum \cite{2021MNRAS.504..926M}, which has a total of 314 GRBs, where 14 GRBs with inaccurate redshifts in the M20 sample have been excluded. 
We further updated the M20 sample with 67 new GRBs until June 2024.
Furthermore, we added three MGFs, GRB 180128A, GRB 200415A, and GRB 231115A, with well-constrained $E_{\rm p}$ from Fermi.
The data of the new GRB samples are listed in Table \ref{table:obs}.
Note that if a parameter has no measured error, we take 10\% of the parameter value as its error.
Finally, we compile a rest frame sample containing 370 GRBs.
In addition, the main emissions (MEs) of GRB 060614, GRB 211211A, GRB 211227A, and GRB 230307A are also included in our sample as different individual events aside from their whole emissions (WEs). 
The ME of GRB 230307A data in the rest frame is collected from \cite{2024ApJ...969...26P}.
GRB 220627A triggered Fermi/GBM twice and was analyzed as two GRBs, which are considered as two individual samples given their different luminous and spectral properties.

Then we estimated the rest frame parameters of the 71 new GRBs with redshift.
The $E_{\rm iso}$ is calculated by
\begin{equation}\label{Eiso}
	E_{\rm iso} = \frac{4\pi D_{\rm L}^2 S_{\gamma}k}{(1+z)},
\end{equation}
where $D_{\rm L}$ is the luminosity distance, $S_{\gamma}$ is the fluence, and $k$ is the $k$--correction factor.
The values of $E_{\rm iso}$ are corrected using $k$--correction to a common rest-frame energy band of 1--$10^{4}$, while simultaneously mitigating the instrumental biases arising from different detector energy coverages.
The $k$--correction factor is defined as
\begin{equation}\label{k}
	k = \frac{{\int_{1/(1 + z)}^{{10^4}/(1 + z)} {EN(E)dE} }}{{\int_{{e_{\min }}}^{{e_{\max }}} {EN(E)dE} }},
\end{equation}
where $e_{\rm min}$ and $e_{\rm max}$ are the observational energy band of fluence, $N(E)$ denotes the photon spectrum of GRB \citep{2007ApJ...660...16S}.
In this work, we assume a flat universe ($\Omega_{\rm k} = 0$) with the cosmological parameters $H_{0}=67.3$ km s$^{-1}$ Mpc$^{-1}$, $\Omega_{\rm M}=0.315$, and $\Omega_{\rm \Lambda}=0.685$, which is consistent with M20.
The rest-frame peak energy $E_{\rm p,z}$ and duration $T_{\rm 90,z}$ are calculated as $E_{\rm p,z} = E_{\rm p}(1 + z)$ and $T_{\rm 90,z} = T_{90}/(1 + z)$, respectively.

\section{The results of t-SNE and UMAP} \label{sec:rest}
\begin{figure*}
	\centering
	\includegraphics[angle=0,scale=0.55]{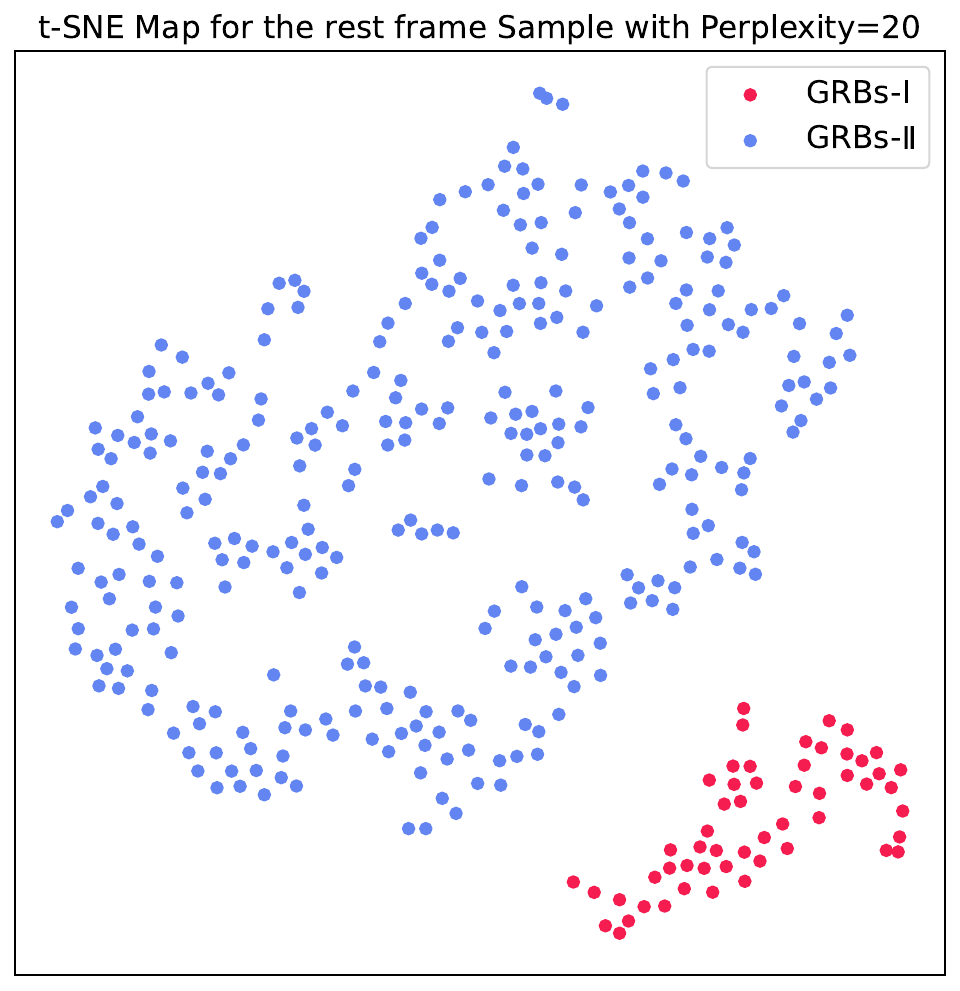}
	\includegraphics[angle=0,scale=0.55]{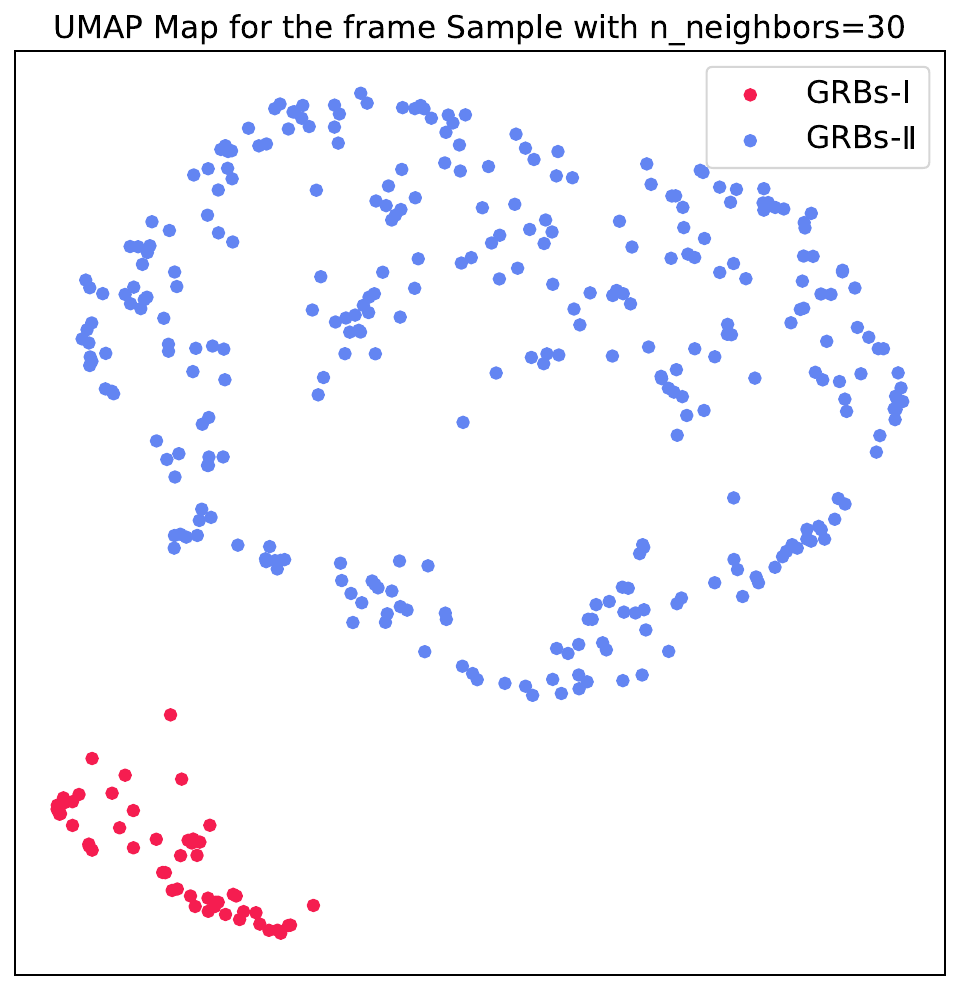}
	\caption{The t-SNE (left) and the UMAP (right) two-dimensional embedding of the 370 GRBs from the rest frame sample based on $T_{\rm 90,z}$, $E_{\rm p,z}$, and $E_{\rm iso}$. There are two clusters: one cluster with dots in red (GRBs-I) and the other cluster with dots in blue (GRBs-II). The axes resulting from t-SNE and UMAP have no clear physical interpretation or units, only the structure is meaningful.}
	\label{f-rest}
\end{figure*}

We obtained t-SNE ($perplexity=20$) and UMAP ($n\_neighbors=30$, $min\_dist=0.01$) maps of GRBs based on both $T_{\rm 90,z}$, $E_{\rm p,z}$, and $E_{\rm iso}$, respectively, as shown in Figure \ref{f-rest}.
The GRBs are clearly divided into two clusters, one is smaller and the other is larger.
To distinguish from the traditional classification methods, we call the small cluster GRBs-I and the larger cluster GRBs-II.
The classification results are listed in Table \ref{table:rest}.

\begin{figure*}
	\centering
	\includegraphics[angle=0,scale=0.38]{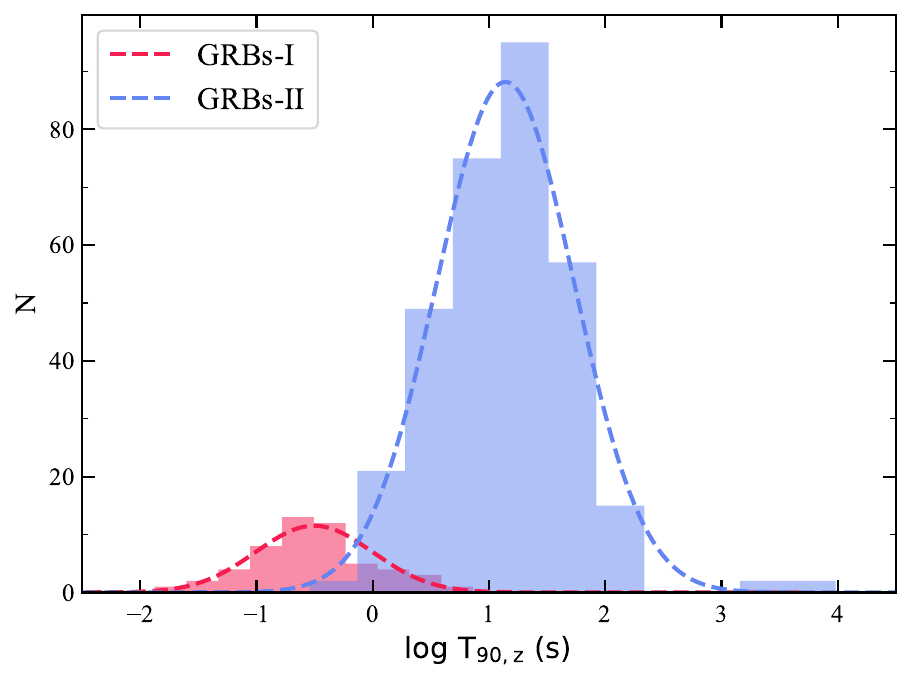}
	\includegraphics[angle=0,scale=0.38]{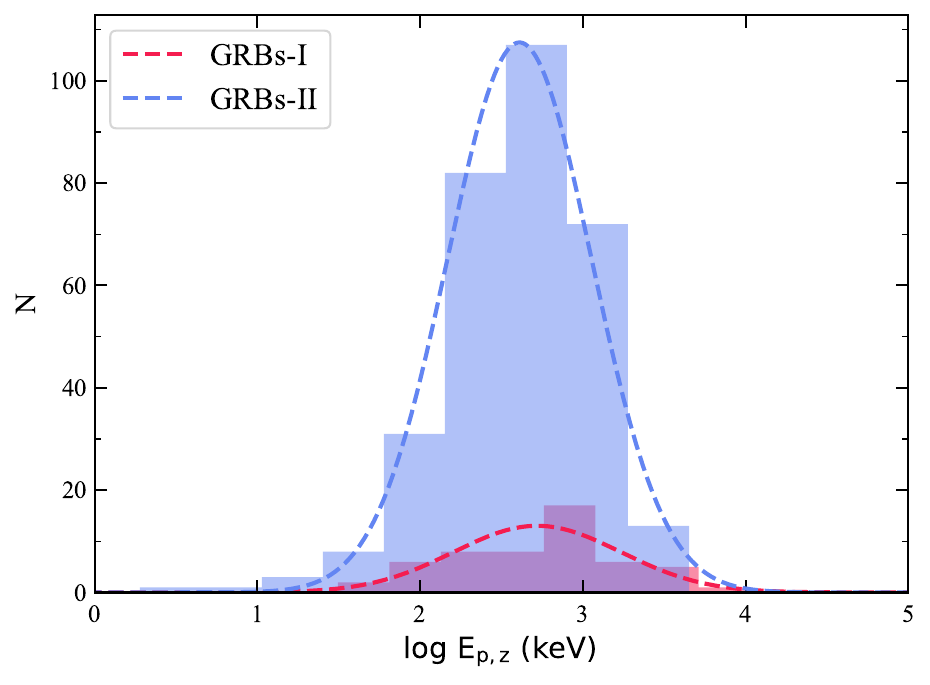}
	\includegraphics[angle=0,scale=0.38]{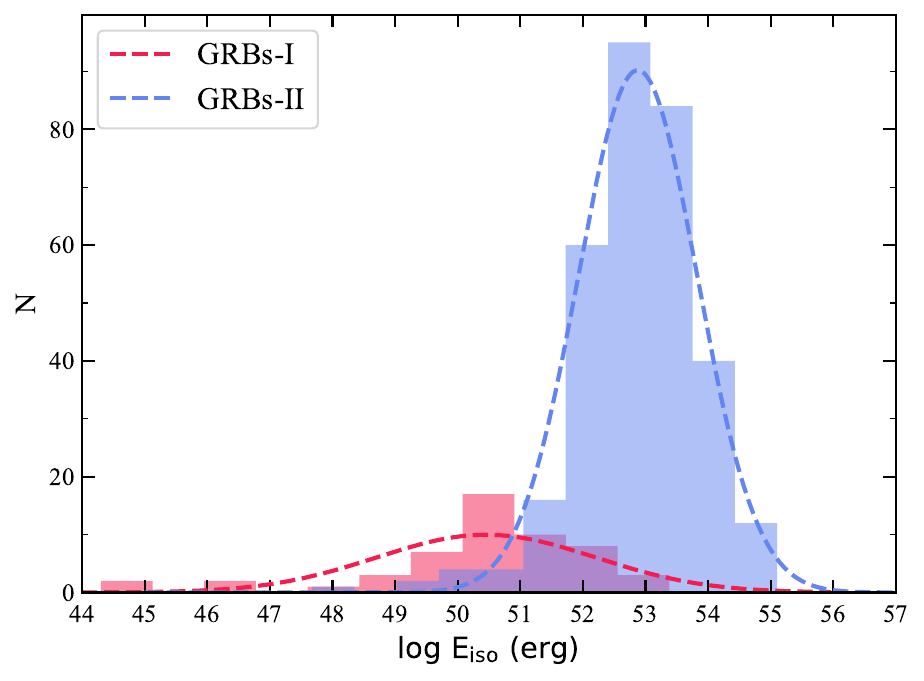}
	\caption{Distributions of $T_{\rm 90,z}$, $E_{\rm p,z}$, and $E_{\rm iso}$ in the rest frame based on UMAP. The dashed lines represent Gaussian fitting curves.}
	\label{f-rest-distributions}
\end{figure*}

We compare the results between t-SNE and UMAP and find that they remarkably agree with each other except for one GRB 110402A, which is classified as GRBs-I on the t-SNE map, while classified as GRBs-II on the UMAP map.
For the t-SNE (UMAP) result, there are 54 (53) GRBs-I, accounting for 14.59\% (14.32\%), and 316 (317) GRBs-II, accounting for 85.41\% (85.68\%).
Since the difference between t-SNE and UMAP is negligible, we only present the distributions of the UMAP result.
The $T_{\rm 90,z}$, $E_{\rm p,z}$, and $E_{\rm iso}$ distributions of the UMAP result are shown in Figure \ref{f-rest-distributions}.
For GRBs-I (GRBs-II), the median values and standard deviations for $T_{\rm 90,z}$, $E_{\rm p,z}$ and $E_{\rm iso}$ are $T_{\rm 90,z} \sim 0.31\ (13.84)$ s and $\sigma \sim 0.50\ (0.59)$, $E_{\rm p,z} \sim 523.83\ (407.94)$ keV and $\sigma \sim 0.51\ (0.44)$, and $E_{\rm iso} \sim 0.28\ (75.19) \times 10^{51}$ erg and $\sigma \sim 1.75\ (0.95)$, respectively.

We find that $T_{\rm 90,z}$ presents a bimodal distribution, and the median value of GRBs-I is significantly smaller than that of GRBs-II.
The $T_{\rm 90,z}$ of GRBs-I can be as long as $\sim 5.33$ s and GRBs-II can be as short as $\sim 0.43$ s.
Furthermore, GRBs-I have a larger $E_{\rm p,z}$ and smaller $E_{\rm iso}$ than GRBs-II, respectively.
Obviously, the parameter distributions of GRBs-I and GRBs-II are similar to SGRBs and LGRBs, respectively \citep{2023ApJ...950...30Z}.

\begin{figure*}
	\centering
	\includegraphics[angle=0,scale=0.36]{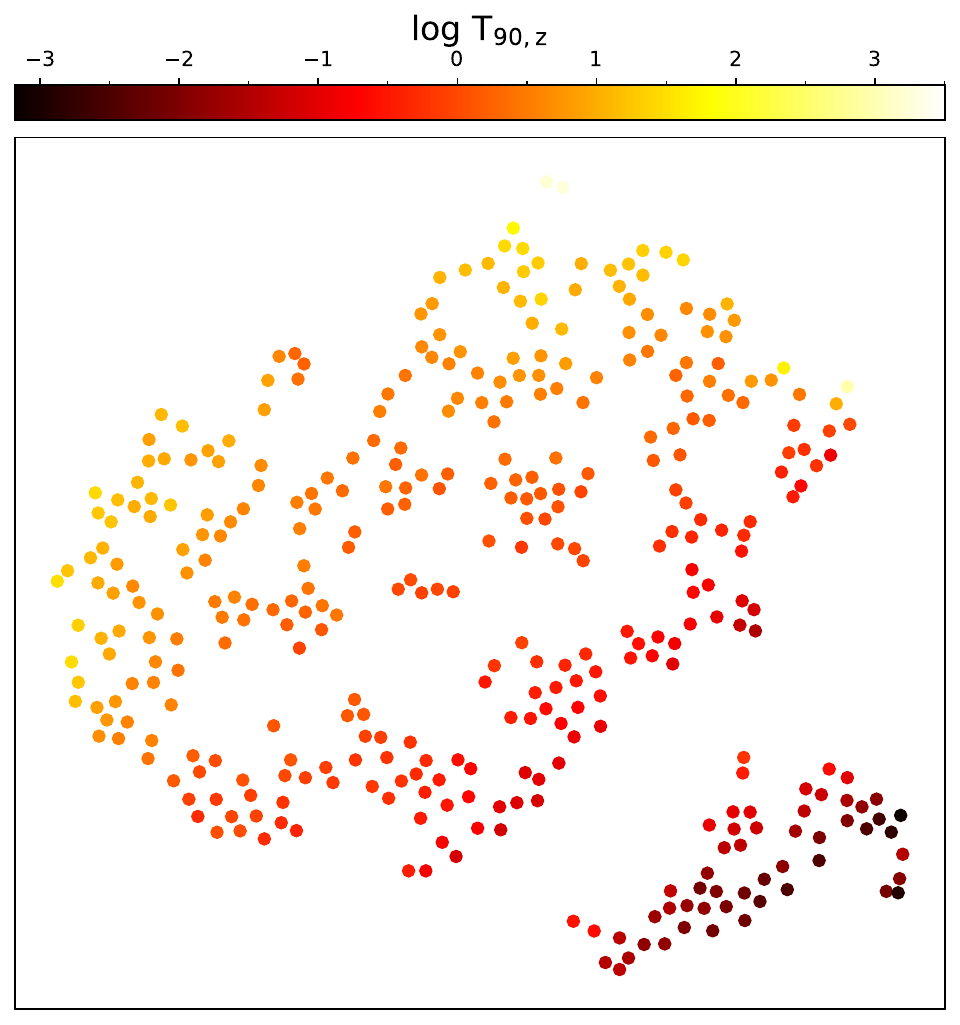}
	\includegraphics[angle=0,scale=0.36]{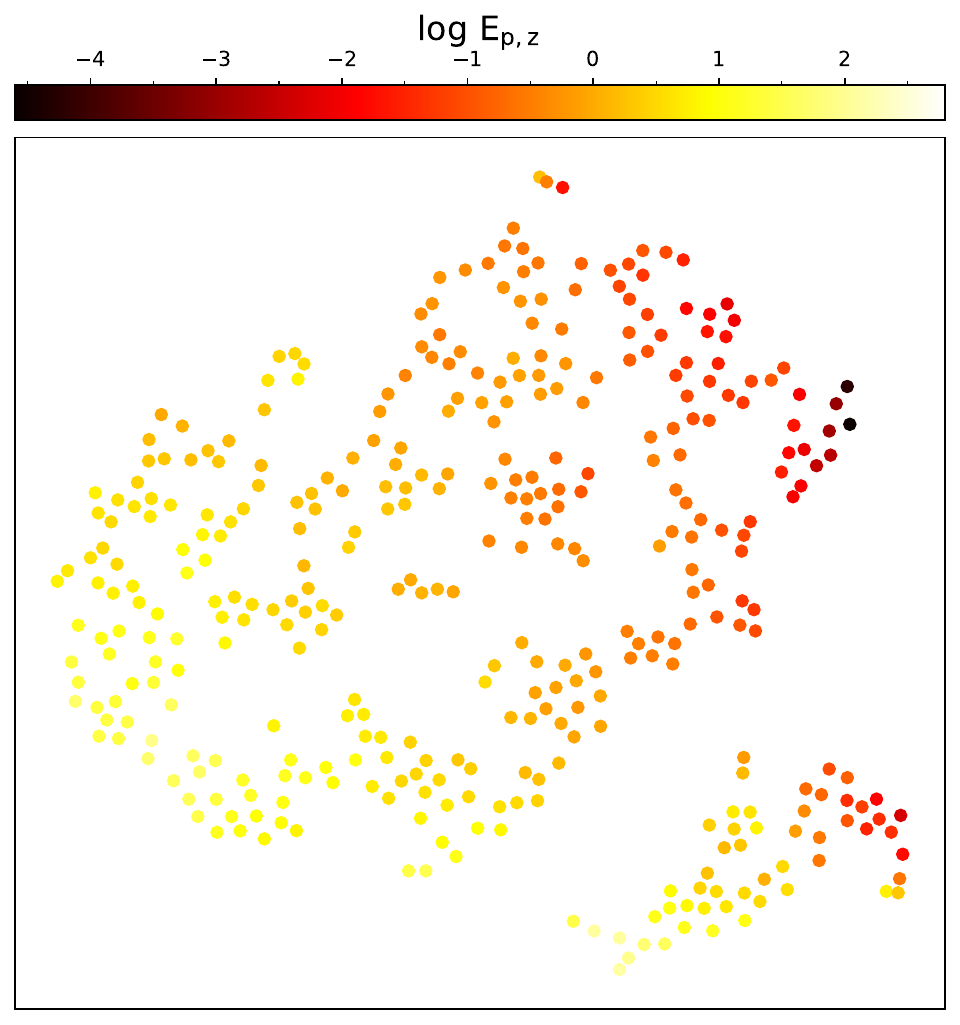}
	\includegraphics[angle=0,scale=0.36]{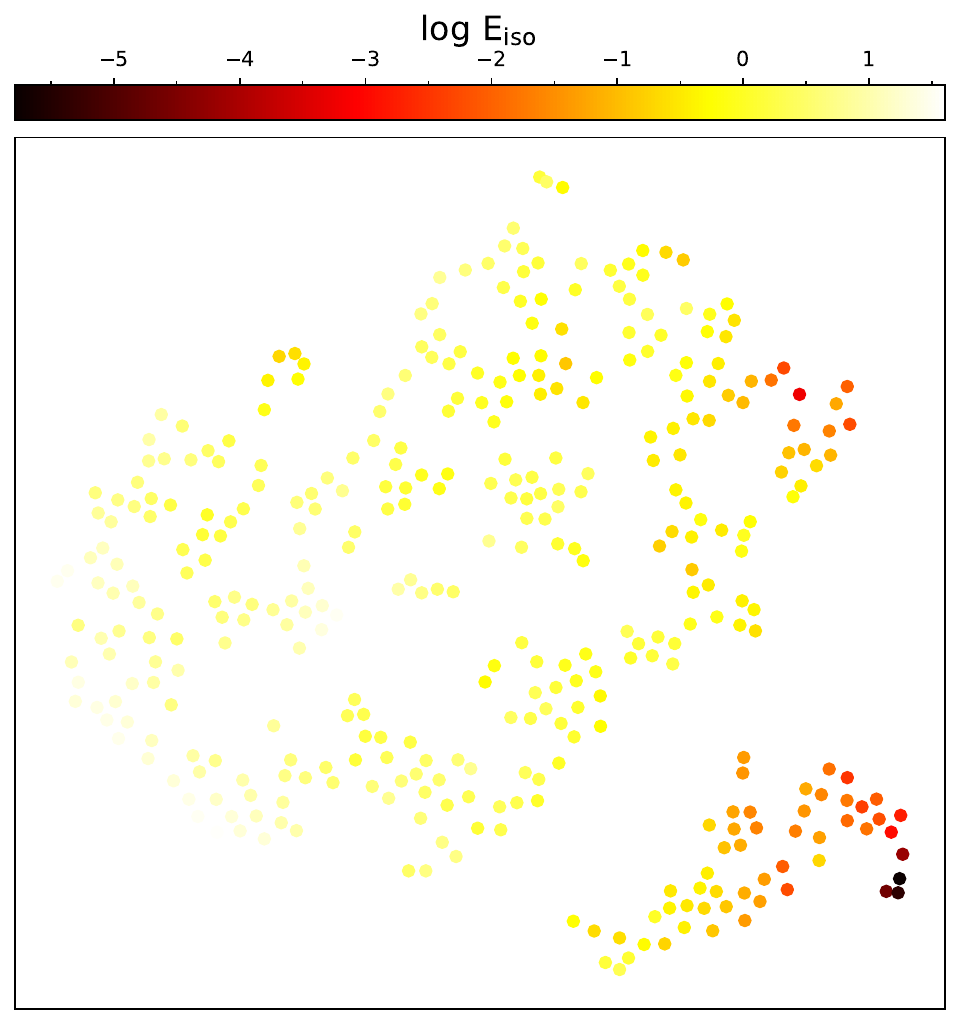}
	\includegraphics[angle=0,scale=0.36]{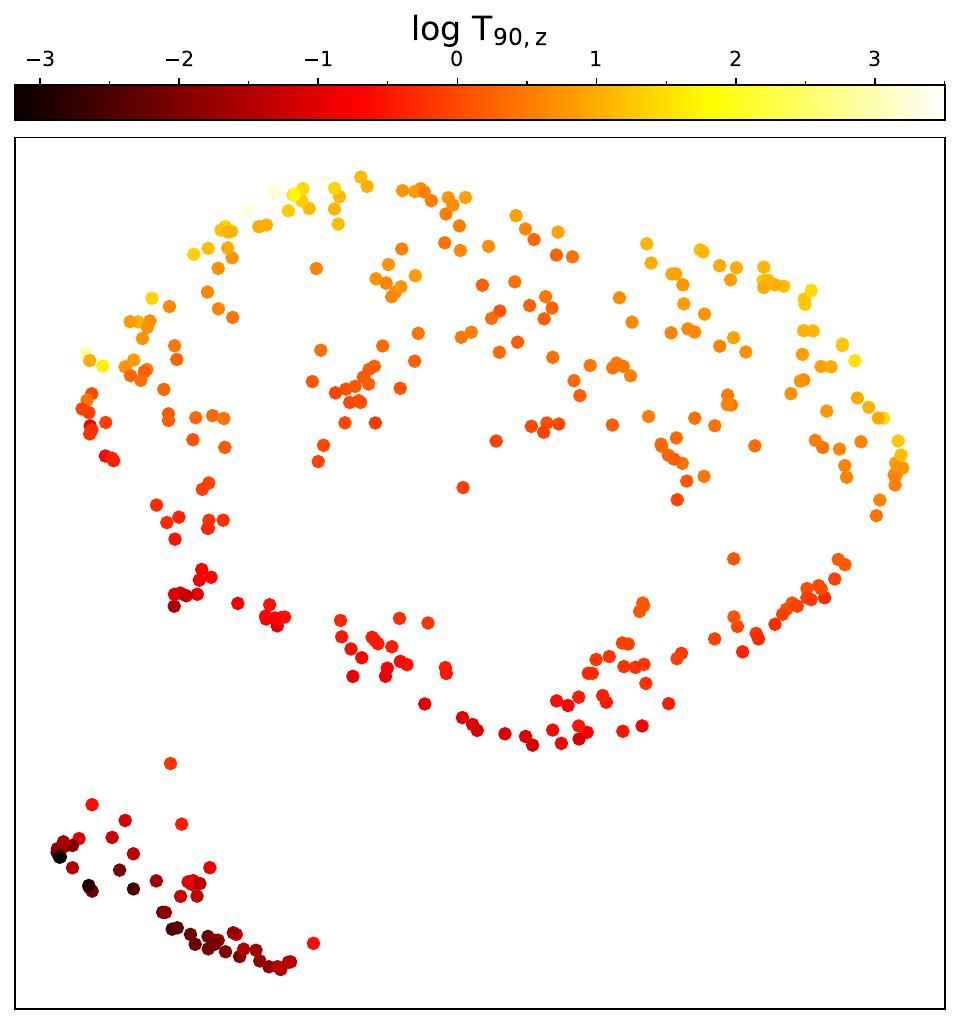}
	\includegraphics[angle=0,scale=0.36]{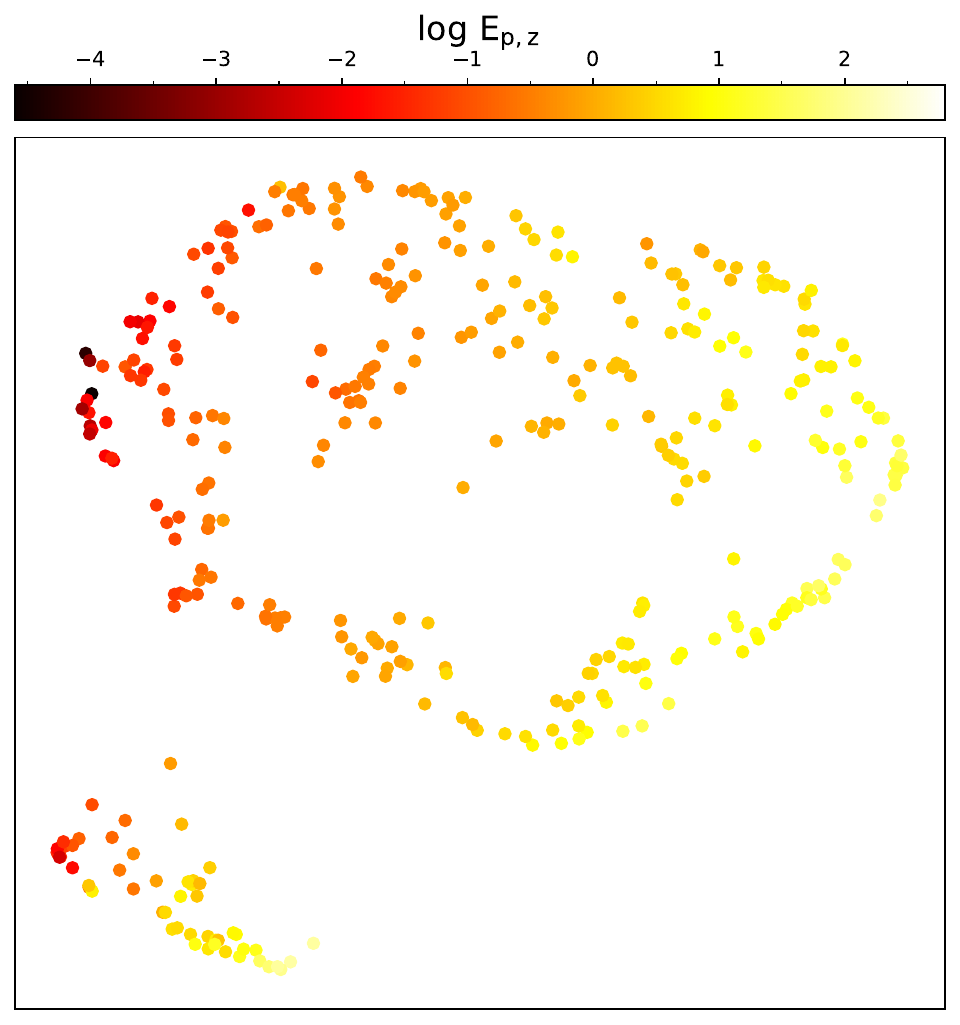}
	\includegraphics[angle=0,scale=0.36]{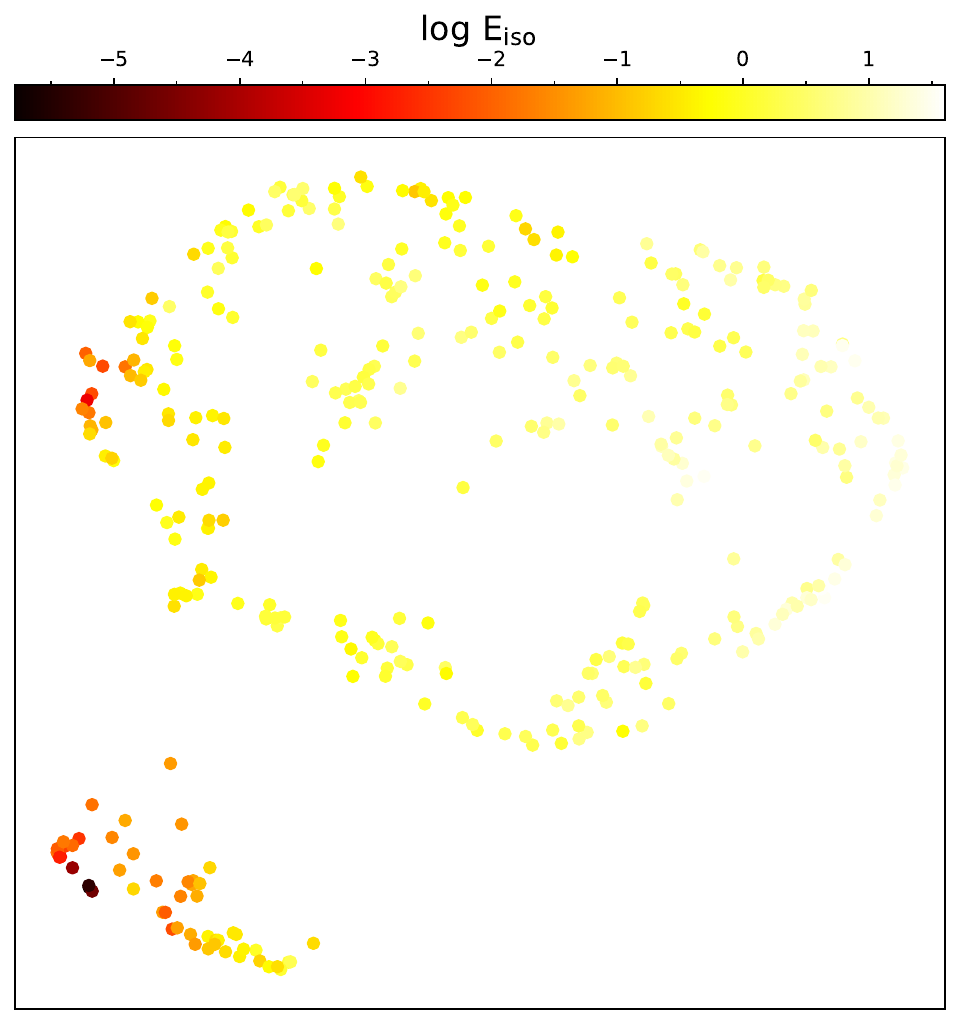}
	\caption{The t-SNE (top) and UMAP (bottom) maps based on the rest frame sample, colored based on $T_{\rm 90,z}$, $E_{\rm p,z}$, and $E_{\rm iso}$, respectively.}
	\label{f-rest-t90z}
\end{figure*}

Moreover, we colored each GRB with rest-frame parameters ($T_{\rm 90,z}$, $E_{\rm p,z}$, and $E_{\rm iso}$) in our t-SNE/UMAP projections.
As shown in Figure \ref{f-rest-t90z}, we observe continuous parameter gradients manifesting as systematic color transitions from dark red (low values) to light yellow (high values).
Interestingly, only the $T_{\rm 90,z}$ exhibits approximately correlated boundaries between GRBs-I and GRBs-II.
This strongly implicates $T_{\rm 90,z}$ as the primary discriminative parameter underlying the classification.
Note that, changing the values of $perplexity$ or $n\_neighbors$ will yield different topological structures, allowing for the study of more refined substructures or macrostructures, without obliterating the traces of GRBs-I and GRBs-II.

\begin{figure}
	\centering
	\includegraphics[angle=0,scale=0.48]{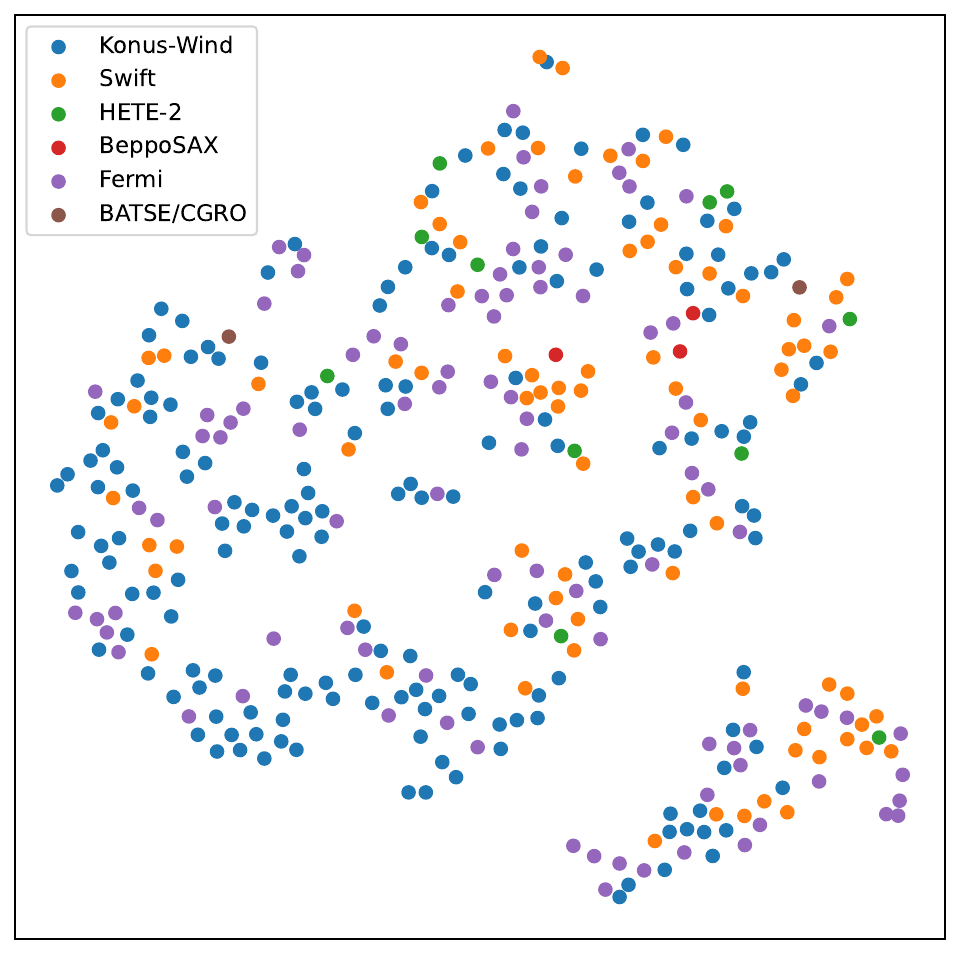}
	\includegraphics[angle=0,scale=0.48]{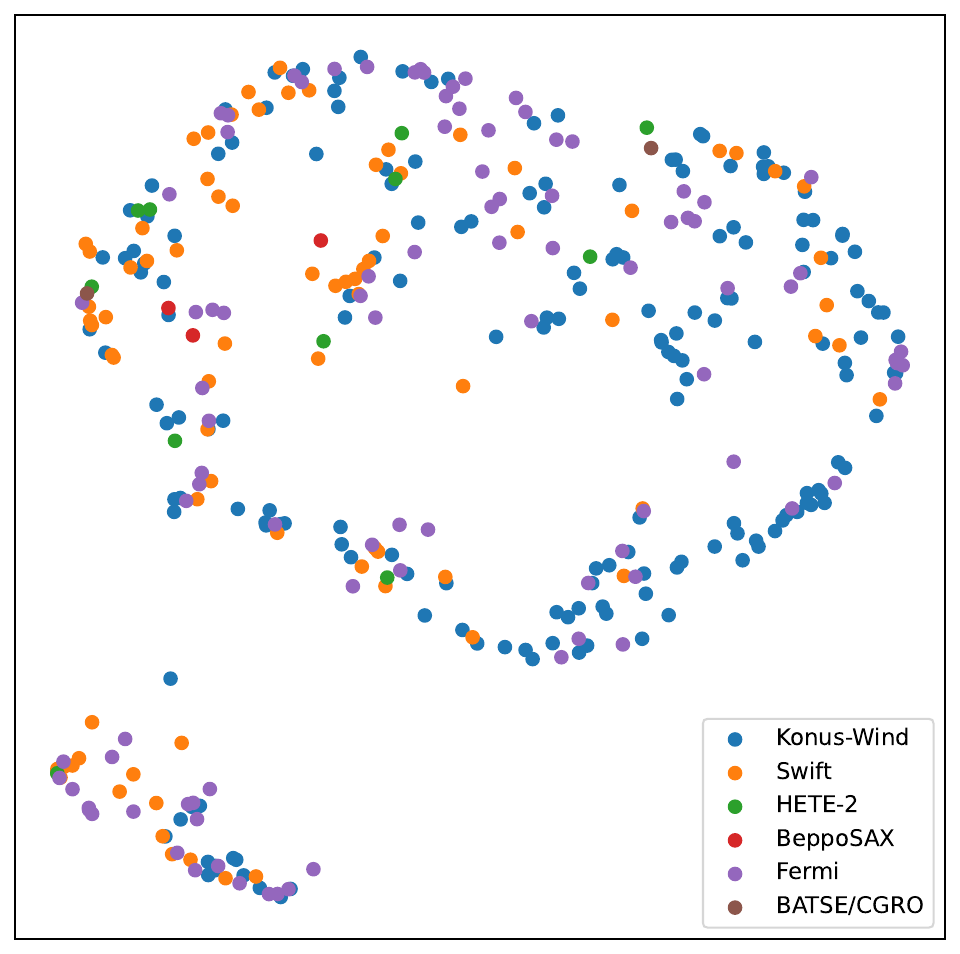}
	\caption{The t-SNE (top) and the UMAP (bottom) maps. The blue, orange, green, red, purple, and brown points represent the prompt emission of GRBs observed by Konus-Wind, Swift, HETE-2, BeppoSAX, Fermi, and BATSE/CGRO, respectively.}
	\label{f-rest-instrument}
\end{figure}

Furthermore, since the GRB samples originate from multiple instruments, this inevitably introduces biases arising from differences in sensitivity and energy coverage across instruments.
We examined the distribution of GRBs observed by different instruments in the t-SNE and UMAP embedding maps. 
As shown in Figure \ref{f-rest-instrument}, GRBs from different instruments do not exhibit obvious clustering in the embedding maps.
In addition, aside from the small samples from HETE-2, BeppoSAX, and BATSE/CGRO, the major contributing instruments (Konus-Wind, Swift, and Fermi) show a reasonable spread across different clusters.
Specifically, 15 Konus-Wind GRBs (28.3\%), 16 Swift GRBs (30.1\%), and 21 Fermi GRBs (39.6\%) are classified as Type I GRBs. These results suggest that instrumental biases do not significantly affect the embedding results and are unlikely to produce artificial structures.

On the other hand, previous studies have shown that $T_{\rm 90,z}$, $E_{\rm p,z}$, and $E_{\rm iso}$ exhibit redshift dependence, i.e., redshift evolution, which may introduce potential biases when such parameters are used for classification \citep{2003A&A...400..415W,2014Ap&SS.350..691Z,2019MNRAS.488.5823L,2020ApJ...904...97D,2022Galax..10...24D,2022ApJ...925...15L,2023ApJ...951...63D}.
Although the nonparametric $\tau$ statistical method has been widely applied to remove redshift evolution from samples, its application generally requires a well-defined flux threshold \citep{1971MNRAS.155...95L,1992ApJ...399..345E}.
\citet{2021ApJ...914L..40D} have reported that the results are not significantly affected by the flux limits. 
In our work, the combination of observations from multiple instruments prevents us from determining a uniform threshold with high precision, which may lead to additional uncertainties when applying the $\tau$ statistical method to correct for redshift evolution.
Given that the influence of redshift evolution on GRB classification remains unclear, we will perform a comprehensive investigation of this issue further in a forthcoming work.

\section{Discussion} \label{sec:discussions}
\subsection{GRBs associated with other electromagnetical counterparts} \label{subsec:counterparts}
\begin{figure*}
	\centering
	\includegraphics[angle=0,scale=0.55]{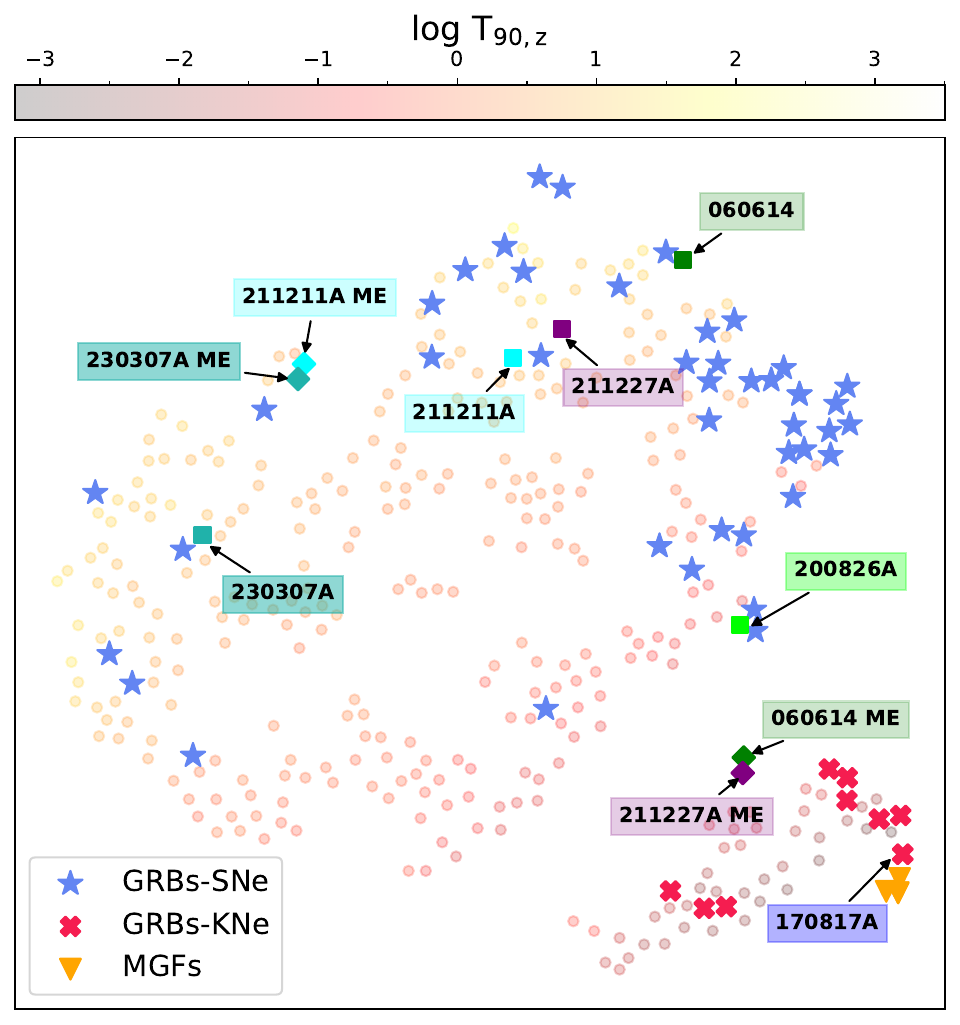}
	\includegraphics[angle=0,scale=0.55]{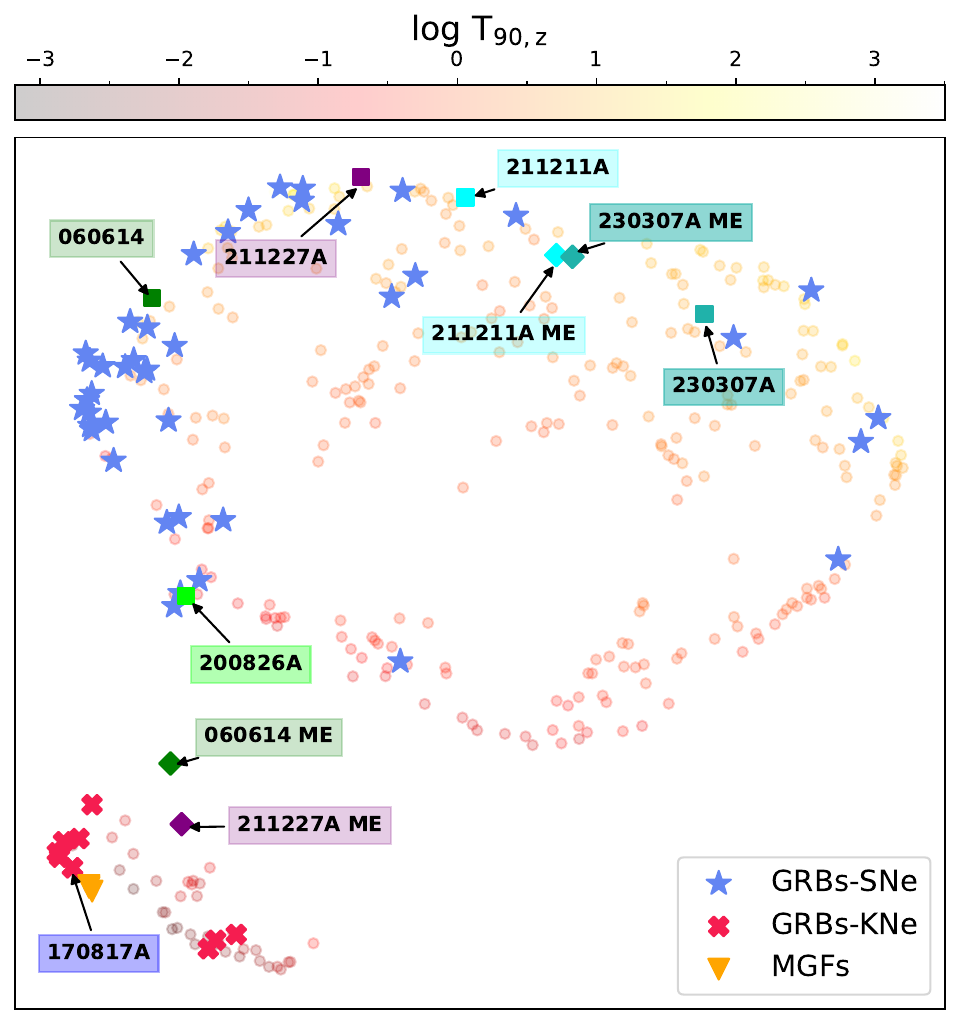}
	\caption{The locations of some special GRBs on the t-SNE and UMAP maps.}
	\label{f-rest-spe}
\end{figure*}

Since unsupervised algorithms do not require prior labeling of samples, the physical nature of these two clusters is unknown.
t-SNE and UMAP indeed cluster GRBs with similar properties, and these two clusters may strongly suggest that they have different origins.
According to whether they are associated with KN/SN, GRBs can be unambiguously classified as merger/collapsar origin.
To investigate the connections between GRBs-I/GRBs-II with merger/collapsar, we located GRBs associated with KNe/SNe and MGFs on the t-SNE and UMAP maps, as shown in Figure \ref{f-rest-spe}.

The rest frame sample includes a total of 43 confirmed GRBs associated with SNe.
They are all classified as GRBs-II, especially for GRB 200826A.
GRB 200826A is a peculiar SGRB with $T_{90}=1.13$ s originating from a collapsar \citep{2021NatAs...5..917A,2021NatAs...5..911Z,2022ApJ...932....1R}.
In addition, the spectral lag of GRB 200826A is 0.157 s, which is at odds for SGRBs but is typical for LGRBs \citep{2006MNRAS.367.1751Y,2017ApJ...844..126S}.
Meanwhile, it is fully consistent with the $E_{\rm p,z}$--$E_{\rm iso}$ correlation of LGRBs \citep{2021NatAs...5..911Z,2023MNRAS.524.1096L}.
Furthermore, we notice that most of GRBs-II associated with SN are concentratedly distributed in the upper right of the t-SNE map and the upper left of the UMAP map, corresponding to the smaller $E_{\rm p,z}$ and $E_{\rm iso}$, respectively.

Meanwhile, nine SGRBs, GRB 050709, GRB 050724A, GRB 061006, GRB 070714B, GRB 070809, GRB 130603B, GRB 150101B, GRB 160821B, and GRB 170817A are associated with KNe, and all are classified as GRBs-I \citep{2013ApJ...774L..23B,2013Natur.500..547T,2016NatCo...712898J,2017ApJ...837...50G,2018NatCo...9.4089T,2019ApJ...883...48L,2019MNRAS.489.2104T,2020NatAs...4...77J}.

In particular, four LGRBs that may be associated with KNe, GRB 060614, GRB 211211A, GRB 211227A, and GRB 230307A are classified as GRBs-II \citep{2015NatCo...6.7323Y,2015ApJ...811L..22J,2022ApJ...931L..23L,2022Natur.612..223R,2022Natur.612..232Y,2022ApJ...936L..10Z,2023A&A...678A.142F,2024Natur.626..737L,2024Natur.626..742Y}.
The light curve of each burst's WE consists of ME and extended emission (EE).
Generally, ME and EE are believed to be powered by different physical mechanisms.
\cite{2022ApJ...936L..10Z} found that GRB 060614, GRB 211211A, and GRB 211227A exhibited unambiguous fallback accretion signatures in their EEs, with mass accretion rate decreasing as $t^{-5/3}$, which supports that EEs are powered by the fallback accretion of r-process heating materials \citep{2019MNRAS.485.4404D}.
\cite{2023ApJ...958L..33G} proposed a unified picture of compact binary mergers via numerical simulations.
They suggested that the merger produced a massive disk that would produce long Type I GRBs, while a light disk would produce short Type I GRBs, and EE arises in the pre-magnetically arrested disk (MAD) phase, with mass accretion rate decreasing as $ t^{-2}$ with time after the transition to MAD.
Therefore, the MEs of GRB 060614, GRB 211211A, GRB 211227A, and GRB 230307A were also analyzed independently of their WEs.

Interestingly, the MEs of GRB 060614 and GRB 211227A are classified as GRBs-I, while the MEs of GRB 211211A and GRB 230307A are classified as GRBs-II.
Although GRB 211211A and GRB 230307A may be associated with KN, their origin remains controversial.
Since theoretical simulations suggest that KN may not be the only explosion in which the r-process occurs, an unusual SN could explain both the duration of GRB 211211A and the r-process-powered excess in its afterglow \citep{2023ApJ...947...55B}.
\cite{2024ApJ...968...14R} also found some GRBs associated with SNe can produce r-process material in special conditions.
If GRB 211211A and GRB 230307A indeed originated from mergers, their MEs were classified as GRBs-II most likely due to a significantly larger $T_{90}$ than the MEs of GRB 060614 and GRB 211227A.
They are outliers in the distribution of GRBs-I and are difficult to be recognized by machine learning.
Note that these two outliers are very close to each other on the t-SNE and UMAP maps, which also suggests that they may have a common origin.

\subsection{Comparison with other classification} \label{subsec:comparison}

\subsubsection{The $T_{90}$ classification} \label{subsubsec:duration}
Firstly, we compare our classification with the traditional $T_{90}$ classification.
We find that there are 52 SGRBs, accounting for 14\%, and 318 LGRBs, accounting for 86\%, in the rest frame sample, which are similar to GRBs-I and GRBs-II, respectively.
However, four SGRBs, GRB 021211, GRB 040924, GRB 150514A, and GRB 200826A are classified as GRBs-II by both t-SNE and UMAP.
Meanwhile, six LGRBs, GRB 050724A, the ME of GRB 060614, GRB 110402A, GRB 161001A, GRB 180618A, and the ME of GRB 211227A are classified as GRBs-I, except for GRB 110402A as mentioned before which is classified as GRBs-II by UMAP.

GRB 021211 and GRB 040924 are associated with SNe, which confirms that they originate from collapsar.
Since the duration of GRB 021211 and GRB 040924 observed by Konus-Wind in our sample are $T_{90}=1.8$ s and $T_{90}=0.8$ s, respectively, they are classified as SGRBs \citep{2017ApJ...850..161T}.
However, their $T_{90}$ observed by HETE-2 are 4.23 s and 3.37 s, respectively, so that they were instead classified as LGRBs \citep{2008A&A...491..157P}.
These differences indicate how unreliable the $T_{90}$ classification is since the measurement of $T_{90}$ strongly depends on the instrument and energy band \citep{2013ApJ...763...15Q}.
Unfortunately, the lack of further observations for GRB 150514A prevents us from determining its natural origin.

GRB 161001A and GRB 180618A are LGRBs with $T_{90}=2.24$ s and $T_{90}=3.71$ s, respectively, while they are classified as GRBs-I.
According to multi-wavelength observations, GRBs originating from mergers usually have a large offset from the center of the host galaxy, small star formation rate (SFR), and small specific star formation rate (sSFR, sSFR = SFR/$M_{\odot}$, where $M_{\odot}$ is the stellar mass of the host galaxy) \citep{2009ApJ...703.1696Z,2010ApJ...708....9F,2020ApJ...897..154L,2022ApJ...940...56F,2022ApJ...940...57N}.
GRB 161001A and GRB 180618A have an offset of $18.54 \pm 6.22$ kpc and $9.7 \pm 1.69$ kpc from their host galaxy, respectively, which is consistent with the merger origin \citep{2022ApJ...940...56F}.
Meanwhile, the host galaxy of GRB 161001A has SFR = $0.53^{+0.59}_{-0.31}$ $M_{\odot}$ yr$^{-1}$ and log(sSFR) = $-10.02^{+0.33}_{-0.37}$ yr$^{-1}$ and the host galaxy of GRB 180618A has SFR = $1.85^{+1.77}_{-1.10}$ $M_{\odot}$ yr$^{-1}$ and log(sSFR) = $-8.54^{+0.44}_{-0.39}$ yr$^{-1}$, which are low-SFR galaxies and consistent with the merger origin \citep{2022ApJ...940...57N}.
Furthermore, GRB 161001A and GRB 180618A both have negligible spectral lags, respectively \citep{2016GCN.19974....1M,2022ApJ...939..106J}.
These results strongly suggest that GRB 161001A and GRB 180618A originated from mergers.

GRB 050724A is actually a SGRB with EE and may be associated with KN, which support that it originated from a merger.
Furthermore, the light curve of ambiguous GRB 110402A from Konus-Wind started with a hard multi-peaked pulse followed, after $\sim 5$ s, by a softer decaying emission, and the total duration is $\sim 70$ s.
The tiny spectral lag for the initial spikes favors a merger origin \citep{2011GCN.11879....1F}.
Unfortunately, the lack of further observations prevents us from conducting in-depth analysis for this event.

\subsubsection{The $E_{\rm p,z}$--$E_{\rm iso}$ correlation classification} \label{subsubsec:amati}

The $E_{\rm p,z}$--$E_{\rm iso}$ correlation has been widely utilized for the classification of GRBs \citep{2013MNRAS.430..163Q,2020MNRAS.492.1919M,2023MNRAS.524.1096L,2023ApJ...950...30Z}.
Although GRBs associated with KNe and those associated with SNe may follow different $E_{\rm p,z}$--$E_{\rm iso}$ correlations, there is overlap between the two GRB populations, while both are distinct from MGFs \citep{2024ApJ...963L..10Y}.
\cite{2020MNRAS.492.1919M} quantitatively classified GRBs into two types based on the $EH$ parameter, Type I with $EH > 3.3$ and Type II with $EH < 3.3$.
To compare the $E_{\rm p,z}$--$E_{\rm iso}$ classification, we present the results of our classification on the $E_{\rm p,z}$--$E_{\rm iso}$ plane, as shown in Figure \ref{f-amati}.

As shown in Table \ref{table:rest}, ten GRBs-I are classified as Type II, as well as five GRBs-II are classified as Type I.
GRB 201221D with $EH=1.35$ is classified as a Type II burst based on the $E_{\rm p,z}$--$E_{\rm iso}$ plane.
However, this burst is a SGRB with $T_{90}=0.14$ s.
The multi-band analysis for GRB 201221D indicates that it originates from a merger \citep{2022MNRAS.516....1D,2022RAA....22g5011Y}.
It has an offset of $29.35 \pm 24.09$ kpc from the host galaxy \citep{2022ApJ...940...56F}.
Meanwhile, its host galaxy has SFR = $2.36^{+0.29}_{-0.26}$ $M_{\odot}$ yr$^{-1}$ and log(sSFR) = $-8.98^{+0.05}_{-0.05}$ yr$^{-1}$ \citep{2022ApJ...940...57N}.
These properties are consistent with the merger origin.
Furthermore, GRB 980425, GRB 031203A, and GRB 171205A are low-luminosity LGRBs associated with SNe. It is generally believed that they originated from collapsars. But those bursts fall into the region of Type I bursts on the $E_{\rm p,z}$--$E_{\rm iso}$ plane. 
Our classification method can well identify those events as GRBs-II/collapsars.

Considering the duration property, \cite{2020MNRAS.492.1919M} further proposed the $EHD$ parameter to classify GRBs, Type I with $EHD > 2.6$ and Type II with $EHD < 2.6$.
Although the accuracy of the $EHD$ classification is improved, GRB 050724A, GRB 060614, GRB 131004A are incorrectly classified as Type II.
Interestingly, our classification can avoid this misjudgment, which indicates that our classification is more effective.

\subsubsection{The prompt--afterglow correlation classification} \label{subsubsec:xray}
According to the standard fireball model of GRBs, the afterglows arise from the interaction between relativistic ejecta from the central engine of GRB and the circumburst medium, producing multi-wavelength non-thermal emission via external shocks. 
However, observations suggest that certain afterglow components, such as the plateau phase, are unlikely to originate from external shocks but instead are probably powered by radiation from an "internal" region within the jet, sustained by the late-time activity of the central engine \citep{2006ApJ...642..354Z}.
For example, the isotropic luminosity ($L_{\rm iso}$) or $E_{\rm iso}$ of prompt emission have tight correlations between the end time of plateau phase ($T_{\rm a,z}$) and the corresponding luminosity at the moment ($L_{\rm X,a}$/$L_{\rm Opt,a}$) in X-ray/optical afterglow  \citep{2011MNRAS.418.2202D,2012A&A...538A.134X,2017ApJ...848...88D,2018ApJ...863...50S,2019ApJS..245....1T}.
These results indicate a close connection between the prompt emission and the plateau phase, and they are likely to have a common origin, i.e., "internal" origin.

\citet{2017ApJ...848...88D} found the SGRBs with EE follow different $L_{\rm X,a}$--$T_{\rm a,z}$--$L_{\rm iso}$ correlations from other LGRBs, potentially constituting a distinct class.
Recently, \citet{2025JHEAp..4700384L} proposed that the plateau phase may also serve as a potential classifier for distinguishing the origins of GRBs.
Inspired by the residual analysis of the correlations performed by \citet{2016ApJ...828...36D} and \citet{2020MNRAS.492.1919M}, they introduced a new classification parameter based on the residuals of the $L_{\rm X,a}$--$T_{\rm a,z}$--$L_{\rm iso}$ correlation, namely plateau shift ($PS$), $PS = {\rm log}(L_{\rm X,a}) + 0.85 \times {\rm log}(T_{\rm a,z}) - 0.68 \times {\rm log}(L_{\rm iso}) - 13.6$.
They proposed that GRBs with $PS < -1$ originate from mergers, whereas those with $PS > -1$ originate from collapsars.

Compared with their results, we find that all GRBs with $PS > -1$ are classified as GRBs-II, whereas typical SGRBs with $PS < -1$ are classified as GRBs-I.
However, for long-duration but intrinsically short (i.e., $T_{\rm 90,z} < 2$ s) GRBs, GRB 100816A ($PS < -1$) and GRB 110731A ($PS < -1$), we classify them as GRBs-II.
The origins of GRB 100816A and GRB 110731A remain controversial.
They have intrinsically short durations and large offsets, suggesting a possible merger origin \citep{2017ApJ...843..114L,2025JHEAp..4700384L}.
In contrast, observations of GRB 200826A indicate that massive collapsars can also produce intrinsic SGRBs, and some LGRBs also have large offset.
\citet{2011ApJ...739...47F} suggested that GRB 100816A is better explained by a wind medium, implying that its progenitor should be a massive star. 
\citet{2012ApJ...750...88Z} calculated its hardness ratio and suggested that it belongs to LGRBs.
\citet{2024MNRAS.532.1434Z} also applied machine learning to Fermi GRBs and found that GRB 100816A and GRB 110731A are classified together with Type II GRBs, which further supports that they originate from collapsars.
Additionally, they follow the same $E_{\rm p,z}$--$E_{\rm iso}$ correlation as Type II GRBs.
In conclusion, they most likely originate from collapsars.

Furthermore, \cite{2023MNRAS.525.5204B} employed unsupervised machine learning algorithms to explore potential links between prompt emission properties and X-ray/optical afterglows.
While afterglow characteristics alone are insufficient to unambiguously determine GRB progenitors, their analysis reveals possible associations between the prompt and afterglow. 
Although a definitive conclusion has yet to be reached, such studies provide a basis for integrating transient and afterglow features into GRB classification, offering a promising direction for future investigations.

\subsection{Magnetar giant flares} \label{subsubsec:mgf}

\begin{figure*}
	\centering
	\includegraphics[angle=0,scale=1]{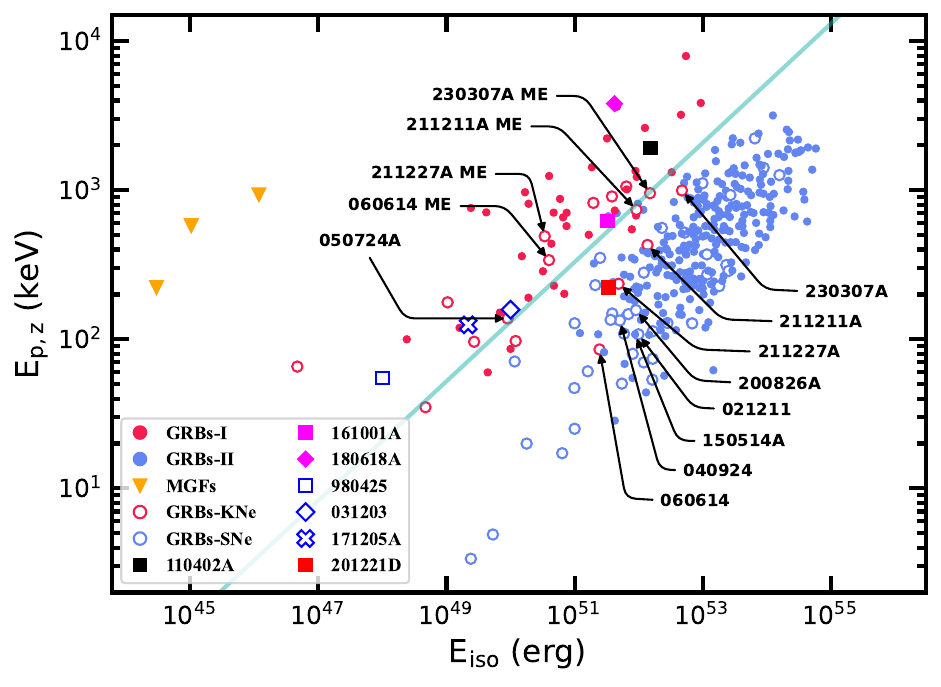}
	\caption{The $E_{\rm p,z}$--$E_{\rm iso}$ plane. The lightseagreen line is the boundary corresponding to $EH = 3.3$.}
	\label{f-amati}
\end{figure*}

GRB 180128A, GRB 200415A, and GRB 231115A are three SGRBs and are considered as MGFs candidates because they are directionally associated with nearby galaxies \citep{2021Natur.589..207R,2021Natur.589..211S,2024A&A...687A.173T,2024Natur.629...58M}.
GRB 180128A and GRB 200415A may be associated with the Sculptor galaxy (NGC 253), and GRB 231115A may be associated with the Cigar galaxy (M82).
Both NGC 253 and M82 are in our local group of galaxies with the luminosity distance $\sim 3.5$ Mpc.
In addition, they are starburst galaxies with high star formation rates, which is contrary to the properties of host galaxies of typical SGRBs \citep{2009ApJ...703.1696Z,2020ApJ...897..154L}.
More importantly, if GRB 231115A had originated from the compact binary merger at 3.5 Mpc, it would have produced a strong signal in gravitational waves, but non-detection has been reported by the LIGO/Virgo/KAGRA Collaboration \citep{2024Natur.629...58M}.
As shown in Figure \ref{f-amati}, GRB 180128A, GRB 200415A, and GRB 231115A are extremely deviate from the classical GRB population, even compared to the off-axis observed GRB 170817A.
However, without accurate localization and multi-band observation, typical SGRBs and MGFs cannot be completely distinguished only from their observed properties of prompt emission \citep{2024MNRAS.532.1434Z}.
Here, three MGFs are all classified as GRBs-I.
Although MGFs are not divided from SGRBs, they are located in adjacent positions on the t-SNE and UMAP maps.
This may also be due to the limited sample size, which was not enough to be divided as a new cluster.

\section{Conclusions} \label{sec:conclusions}
We employ dimensionality reduction algorithms t-SNE and UMAP to classify GRBs using the rest-frame prompt emission parameters ($T_{\rm 90,z}$, $E_{\rm p,z}$, and $E_{\rm iso}$).
Both algorithms consistently resolve the GRB population into two distinct clusters: GRBs-I and GRBs-II.
Notably, while the $T_{\rm 90,z}$ distributions of both classes exhibit significant overlap, GRBs-I demonstrate extended durations up to $\sim 5.33$ s, whereas GRBs-II exhibit durations as brief as $\sim 0.43$ s.

All SGRBs associated with KNe are classified as GRBs-I.
Conversely, all GRBs associated with SNe, including the enigmatic short-duration burst GRB 200826A (a confirmed collapsar event), resides within GRBs-II. 
Crucially, two long-duration GRBs, GRB 060614 and GRB 211227A, which may originate from mergers, are also classified as GRBs-I.
This systematic clustering strongly supports the hypothesis that GRBs-I predominantly originate from compact binary mergers, while GRBs-II stem from massive collapsars.
Interestingly, the extreme events GRB 211211A and GRB 230307A, if confirmed as merger products, present an anomaly: despite their merger origin, their MEs of prompt emission characteristics align with GRBs-II.
This discrepancy may arise from their exceptionally prolonged $T_{90}$ values compared to the MEs of GRB 060614 and GRB 211227A.
Such outliers suggest the potential existence of a transitional GRB subclass bridging merger and collapsar populations, possibly governed by distinct progenitor physics or observational selection effects.

Our methodology demonstrates superior discriminative capability relative to conventional $T_{90}$-based classification.
Critically, it successfully categorizes ambiguous cases: the collapsar-origin SGRBs (GRB 021211, GRB 040924, GRB 200826A) are reclassified as GRBs-II. Merger-driven LGRBs (GRB 050724A, the ME of GRB 060614, GRB 161001A, GRB 180618A, and the ME of GRB 211227A) are distinctly identified as GRBs-I. Notably, low-luminosity collapsar LGRBs (GRB 980425, GRB 031203, GRB 171205A) consistently cluster within GRBs-II, while the merger-associated short-duration burst GRB 201221D aligns with GRBs-I.
These classifications rectify systematic misidentifications prevalent in $E_{\rm p,z}$--$E_{\rm iso}$ correlation methods, confirming our approach's enhanced capability in grouping bursts by progenitor type.

The implementation of t-SNE and UMAP for rest-frame parameter visualization establishes a new paradigm in GRB classification.
Unlike traditional methods constrained by single-instrument biases or isolated parameter spaces ($T_{90}$ or $E_{\rm p,z}$--$E_{\rm iso}$ correlation), our multidimensional framework simultaneously resolves both classification challenges through intrinsically physical parameter synthesis.
The launch of the Space-based multi-band Variable Objects Monitor (SVOM) satellite and Einstein Probe (EP) will deliver enhanced GRB catalogs with improved redshift completeness.
The application of our rest-frame clustering methodology to these datasets is anticipated to establish robust progenitor-classification benchmarks and optimize follow-up observation strategies through machine learning-predicted progenitor types.

\begin{acknowledgements}
We thank the anonymous reviewers for their insightful comments/suggestions.
We acknowledge the use of public data from the GCN Circulars Archive. 
This work was supported in part by the National Natural Science Foundation of China (No. 12463008), and by the Guangxi Natural Science Foundation (No. 2022GXNSFDA035083).
S.-Y.Z. and P.-H.T.T. thank the support from the National Natural Science Foundation of China (No. 12273122), National Astronomical Data Center, the Greater Bay Area, under grant No. 2024B1212080003, and science research grant from the China Manned Space Project under CMS-CSST-2025-A13.
\end{acknowledgements}

\bibliography{ref}{}
\bibliographystyle{aa}

\begin{appendix}

\FloatBarrier
\onecolumn
\section{Tables}
We list the prompt emission parameters of the 70 newly added GRBs in the observer frame in Table A.1, and the prompt emission parameters and classification results of 370 GRBs in the rest frame in Table A.2, for use in the analyses presented in the main text.
	\begin{longtable}{lccccccccccc}
	
	\caption{The prompt emission parameters of GRBs in the observer frame}\\
	\label{table:obs}
	\setlength{\tabcolsep}{10pt}
	$GRB$ & z & $T_{90}$ & model & $E_{\rm p}$ & $\alpha$ & $\beta$ & $S_{\gamma,6}$ & $e_{\rm min}-e_{\rm max}$ & Ref. \\
	  &  & (s) &  & (keV) &  &  & (erg cm$^{-2}$) & (keV) & \\
	\hline
	\endfirsthead
	\\
	\caption{continued.}\\
	\hline\hline
	$GRB$ & z & $T_{90}$ & Model & $E_{\rm p}$ & $\alpha$ & $\beta$ & $S_{\gamma,6}$ & $e_{\rm min}-e_{\rm max}$ & Ref. \\
	  &  & (s) &  & (keV) &  &  & (erg cm$^{-2}$) & (keV) & \\
	\hline
	\endhead
	\hline
	\endfoot
	060614\_ME & 0.125     & 6$\pm$0.6 & CPL       & 302$^{+214}_{-85}$ & -1.57     & --- & 8.19$\pm$0.56 & 20-20000  & (1) \\
	060614   & 0.125     & 123.65$\pm$18.95 & CPL      & 76$^{+20}_{-29}$ & -1.92      & --- & 47$\pm$2.6 & 10-10000   & (2) \\
	101224A   & 0.454     & 1.73$\pm$1.68 & CPL       & 558.48$^{+270.93}_{-270.93}$ & -1.03     &---& 0.28$\pm$0.06 & 10-1000   & (3) \\
	130716A   & 2.2       & 0.77$\pm$0.39 & CPL       & 818.94$^{+381.37}_{-381.37}$ & -0.48     &---& 0.74$\pm$0.09 & 10-1000   & (3) \\
	140930B   & 1.465     & 1$\pm$0.1 & CPL       & 1302$^{+2009}_{-459}$ & -0.6      &---& 8.1$\pm$5.1 & 20-15000  & GCN 16868 \\
	150831A   & 1.18      & 1$\pm$0.1 & CPL       & 564$^{+122}_{-122}$ & -0.5      &---& 2.4$\pm$0.4 & 20-10000  & GCN 18226 \\
	161001A   & 0.67      & 2.24$\pm$0.23 & CPL       & 373.17$^{+58.58}_{-58.58}$ & -0.94     &---& 2.36$\pm$0.19 & 10-1000   & (3) \\
	170728B   & 1.272     & 46.34$\pm$0.81 & CPL       & 159.05$^{+16.03}_{-16.03}$ & -1.04     &---& 3.78$\pm$0.22 & 10-1000   & (3) \\
	180128A   & 0.00078   & 0.21$\pm$0.4 & CPL       & 222.95$^{+28}_{-28}$ & 1.44      &---& 0.21$\pm$0.03 & 10-1000   & (3) \\
	180314A   & 1.445     & 28$\pm$2.8 & CPL       & 104$^{+3}_{-3}$ & -0.36     &---& 15.2$\pm$0.3 & 10-1000   & GCN 23320 \\
	180618A   & 0.52      & 3.71$\pm$0.58 & CPL       & 2507.58$^{+917.74}_{-917.74}$ & -1.13     &---& 2.15$\pm$0.14 & 10-1000   & (3) \\
	180703A   & 0.6678    & 20.74$\pm$1.56 & Band      & 350.79$^{+32.25}_{-32.25}$ & -0.78     & -1.97     & 19.9$\pm$0.41 & 10-1000   & (3) \\
	180805B   & 0.6612    & 0.96$\pm$0.59 & CPL       & 346$^{+75}_{-75}$ & -0.5      &---& 0.59$\pm$0.07 & 10-1000   & GCN 23078 \\
	181010A   & 1.39      & 60$\pm$6  & CPL       & 280$^{+80}_{-80}$ & -0.8      &---& 0.61$\pm$0.1 & 10-1000   & GCN 22485 \\
	190324A   & 1.1715    & 26.9$\pm$2.69 & Band      & 144.3$^{+8.4}_{-8.4}$ & -0.84     & -2.06     & 17.8$\pm$0.3 & 10-1000   & GCN 24002 \\
	190613A   & 2.78      & 18$\pm$1.8 & CPL       & 108$^{+5}_{-5}$ & 0.06      &---& 3.09$\pm$0.13 & 10-1000   & GCN 24816 \\
	190719C   & 2.469     & 175$\pm$17.5 & CPL       & 81$^{+9}_{-9}$ & -0.87     &---& 1.16$\pm$0.08 & 10-1000   & GCN 25130 \\
	191004B   & 3.503     & 5.5$\pm$0.55 & CPL       & 172$^{+27}_{-21}$ & -0.51     &---& 4.13$\pm$0.45 & 20-10000  & GCN 25974 \\
	191011A   & 1.722     & 13$\pm$1.3 & CPL       & 89$^{+23}_{-23}$ & -1.24     &---& 0.43$\pm$0.05 & 10-1000   & GCN 26000 \\
	191221B   & 1.148     & 35$\pm$3.5 & Band      & 377$^{+30}_{-29}$ & -0.81     & -2.47     & 100$\pm$10 & 20-20000  & GCN 26576 \\
	200415A   & 0.00078   & 0.14$\pm$0.04 & CPL       & 934.13$^{+49.2}_{-49.2}$ & 0.01      &---& 5.15$\pm$0.11 & 10-1000   & (3) \\
	200524A   & 1.256     & 37.76$\pm$6.21 & Band      & 192$^{+16}_{-16}$ & -0.66     & -1.77     & 20.8$\pm$0.3 & 10-1000   & GCN 27809 \\
	200613A   & 1.22      & 478.03$\pm$3.17 & Band      & 111$^{+3}_{-3}$ & -1.08     & -2.58     & 41$\pm$0.5 & 10-1000   & GCN 27930 \\
	200826A   & 0.7481    & 1.14$\pm$0.13 & Band      & 89.8$^{+3.7}_{-3.7}$ & -0.41     & -2.4      & 4.8$\pm$0.1 & 10-1000   & (3) \\
	201015A   & 0.426     & 9.78$\pm$3.47 & Band      & 14$^{+6}_{-6}$ & -1        & -2.4      & 0.23$\pm$0.04 & 10-1000   & GCN 28663 \\
	201020B   & 0.804     & 15.87$\pm$0.36 & Band      & 136.9$^{+3.4}_{-3.4}$ & -0.71     & -2.2      & 39.3$\pm$0.4 & 10-1000   & GCN 28710 \\
	201021C   & 1.07      & 35.33$\pm$2.2 & CPL       & 131.1$^{+28.2}_{-28.2}$ & -1.1      &---& 1.7$\pm$0.2 & 10-1000   & GCN 28748 \\
	201103B   & 1.105     & 105$\pm$10.5 & Band      & 403$^{+74}_{-60}$ & -0.72     & -2.54     & 52.6$\pm$8.2 & 10-10000  & GCN 28872 \\
	201216C   & 1.1       & 29.95$\pm$0.57 & Band      & 326$^{+7}_{-7}$ & -1.06     & -2.25     & 141$\pm$6 & 10-10000  & GCN 29073 \\
	201221A   & 5.7       & 44.5$\pm$6.2 & Band      & 194$^{+81}_{-63}$ & -0.95     & -2.39     & 4.3$\pm$0.7 & 15-1500   & GCN 29150 \\
	201221D   & 1.055     & 0.14$\pm$0.07 & CPL       & 108$^{+5}_{-5}$ & -0.2      &---& 1.08$\pm$0.05 & 10-1000   & GCN 29140 \\
	210104A   & 0.46      & 35$\pm$3.5 & Band      & 157$^{+23}_{-18}$ & -1        & -2.52     & 21.5$\pm$3.5 & 20-10000  & GCN 29258 \\
	210204A   & 0.876     & 206.85$\pm$2.29 & CPL       & 162$^{+33}_{-33}$ & -1.51     &---& 5.76$\pm$0.04 & 10-1000   & GCN 29393 \\
	210210A   & 0.715     & 8.8$\pm$0.88 & CPL       & 16.6$^{+7.2}_{-10.7}$ & -1.68     &---& 1.2$\pm$0.1 & 15-1500   & GCN 29517 \\
	210323A   & 0.733     & 0.96$\pm$0.78 & CPL       & 2100$^{+400}_{-400}$ & -0.97     &---& 1.2$\pm$0.1 & 10-1000   & GCN 29709 \\
	210610A   & 3.54      & 8.19$\pm$2.06 & CPL       & 247$^{+74}_{-74}$ & -1.11     &---& 1.78$\pm$0.23 & 10-1000   & GCN 30233 \\
	210610B   & 1.13      & 55.04$\pm$0.72 & CPL       & 414.3$^{+11.7}_{-11.7}$ & -0.28     &---& 17.3$\pm$0.3 & 10-1000   & GCN 30199 \\
	210619B   & 1.937     & 54.79$\pm$0.57 & Band      & 210$^{+3}_{-3}$ & -0.86     & -1.99     & 308$\pm$0.98 & 10-1000   & GCN 30279 \\
	210702A   & 1.16      & 90$\pm$9  & CPL       & 376$^{+63}_{-61}$ & -0.91     & -1.91     & 2.5$\pm$20 & 20-20000  & GCN 30366 \\
	210722A   & 1.145     & 61.95$\pm$11.34 & CPL       & 160$^{+30}_{-30}$ & -1.1      &---& 5.8$\pm$0.6 & 10-1000   & GCN 30490 \\
	210731A   & 1.2525    & 25.86$\pm$5.28 & CPL       & 175$^{+11}_{-11}$ & -0.1      &---& 4.9$\pm$0.2 & 10-1000   & GCN 30573 \\
	210822A   & 1.736     & 12$\pm$1.2 & Band      & 398$^{+30}_{-29}$ & -0.64     & -2.39     & 120$\pm$11 & 20-20000  & GCN 30694 \\
	211023A   & 0.39      & 79.11$\pm$0.57 & Band      & 92$^{+2}_{-2}$ & -1.74     & -2.55     & 111$\pm$0.7 & 10-1000   & GCN 30965 \\
	211211A\_ME & 0.076     & 13$\pm$1.3 & Band      & 687.1$^{+12.55}_{-11}$ & -1        & -2.36     & 377$\pm$1 & 10-1000   & (1) \\
	211211A   & 0.076     & 43.18$\pm$0.06 & Band      & 399.3$^{+14}_{-16.1}$ & -1.2      & -2.05     & 542$\pm$8 & 10-1000   & (1) \\
	211227A\_ME & 0.228     & 4$\pm$0.4 & CPL       & 400$^{+1200}_{-200}$ & -1.56     &---& 2.01$\pm$0.19 & 15-1500   & (4) \\
	211227A   & 0.228     & 82.5$\pm$8.25 & Band      & 192$^{+45}_{-42}$ & -1.34     & -2.26     & 26$\pm$2.1 & 15-1500   & GCN 31544 \\
	220101A   & 4.618     & 128.26$\pm$15.79 & CPL       & 330$^{+15}_{-15}$ & -1.09     &---& 77$\pm$1  & 10-1000   & GCN 31360 \\
	220107A   & 1.246     & 33.03$\pm$0.57 & Band      & 168$^{+11}_{-11}$ & -0.55     & -1.94     & 22.4$\pm$0.5 & 10-1000   & GCN 31406 \\
	220117A   & 4.961     & 40$\pm$4  & Band      & 65$^{+20}_{-14}$ & -1.04     & -2.59     & 2.8$\pm$0.7 & 15-1500   & GCN 31511 \\
	220219B   & 0.293     & 47$\pm$4.7 & CPL       & 39$^{+27}_{-27}$ & -1.86     &---& 12.1$\pm$1.6 & 20-10000  & GCN 31646 \\
	220521A   & 5.6       & 13.57$\pm$3.9 & CPL       & 37$^{+7}_{-7}$ & -1.51     &---& 0.66$\pm$0.05 & 10-1000   & GCN 32089 \\
	220527A   & 0.857     & 10.5$\pm$0.36 & Band      & 151.6$^{+2.7}_{-2.7}$ & -0.75     & -2.55     & 56.9$\pm$0.4 & 10-1000   & GCN 32133 \\
	220627A\_1 & 3.084     & 136.71$\pm$1.28 & CPL       & 410$^{+16}_{-16}$ & -0.83     &---& 78.8$\pm$1.3 & 10-1000   & GCN 32288 \\
	220627A\_2 & 3.084     & 126.98$\pm$7.67 & CPL       & 227$^{+15}_{-15}$ & -1.12     &---& 24.7$\pm$0.8 & 10-1000   & GCN 32288 \\
	221226B   & 2.694     & 4.86$\pm$2.83 & CPL       & 104$^{+8}_{-8}$ & 0.2       &---& 0.78$\pm$0.05 & 10-1000   & GCN 33112 \\
	230204B   & 2.142     & 216.07$\pm$0.57 & CPL       & 783$^{+40}_{-40}$ & -0.97     &---& 139$\pm$1.5 & 10-1000   & GCN 33288 \\
	230307A   & 0.065     & 34.56$\pm$0.57 & CPL       & 936$^{+3}_{-3}$ & -1.07     &---& 2950$\pm$4 & 10-1000   & (3) \\
	230812B   & 0.36      & 3.26$\pm$0.09 & Band      & 273$^{+3}_{-3}$ & -0.8      & -2.47     & 252$\pm$0.02 & 10-1000   & GCN 34391 \\
	230818A   & 2.42      & 9.98$\pm$0.64 & CPL       & 260$^{+30}_{-30}$ & -1.02     &---& 5.6$\pm$0.3 & 10-1000   & GCN 34501 \\
	231115A   & 0.00078   & 0.03$\pm$0.04 & CPL       & 580$^{+60}_{-60}$ & 0.5       &---& 0.63$\pm$0.04 & 10-1000   & (3) \\
	231117A   & 0.257     & 0.7$\pm$0.07 & Band      & 400$^{+89}_{-82}$ & -1.15     & -2.81     & 9.1$\pm$0.97 & 20-10000  & GCN 35079 \\
	231118A   & 0.8304    & 5.76$\pm$0.57 & Band      & 217.7$^{+22.6}_{-22.6}$ & -0.77     & -2.17     & 10$\pm$1  & 10-1000   & GCN 35131 \\
	231210B   & 3.13      & 9$\pm$0.9 & CPL       & 174$^{+60}_{-27}$ & -0.62     &---& 4.02$\pm$0.56 & 20-15000  & GCN 35359 \\
	240205B   & 0.824     & 47$\pm$4.7 & Band      & 34$^{+0.8}_{-0.8}$ & -1.69     & -2.78     & 47.7$\pm$0.5 & 10-1000   & GCN 35693 \\
	240218A   & 6.782     & 38$\pm$3.5 & Band      & 100$^{+20}_{-20}$ & -0.3      & -1.9      & 5.2$\pm$0.4 & 10-1000   & GCN 35755 \\
	240225B   & 0.946     & 35$\pm$3.5 & Band      & 270$^{+64}_{-63}$ & -1.37     & -2.74     & 21.6$\pm$3.5 & 20-10000  & GCN 35835 \\
	240315C   & 4.859     & 47$\pm$4.7 & CPL       & 410$^{+388}_{-159}$ & -1.09     &---& 16.3$\pm$6.4 & 20-10000  & GCN 35792 \\
	240402B   & 1.551     & 5.1$\pm$0.51 & Band      & 86$^{+5}_{-4}$ & -0.71     & -3.58     & 7.56$\pm$0.5 & 20-10000  & GCN 36028 \\
	240414A   & 1.833     & 82$\pm$8.2 & CPL       & 150$^{+20}_{-20}$ & -0.9      &---& 5.56$\pm$1.07 & 10-1000   & GCN 36120 \\
	\end{longtable}
	\tablebib{(1) \citet{2022Natur.612..232Y}; (2) \citet{2017ApJ...850..161T}; (3) \citet{2020ApJ...893...46V}; (4) \citet{2022ApJ...936L..10Z}.
	}

\begin{table}[h!]
	\caption{The prompt emission parameters of GRBs of the rest sample}
	\label{table:rest}
	\centering                        
	\begin{tabular}{lccccccccccc}      
		\hline\hline               
		$GRB$ & $T_{90}$ & S/L$^{a}$ & $z$ & $T_{\rm 90,z}$ & $E_{\rm p,z}$ & $E_{\rm iso,50}$ & $EH^{b}$ & $EHD^{b}$ & t-SNE$^{c}$ & UMAP$^{c}$ & Association\\
		\hline
		970228 & 53.39     & L         & 0.695     & 31.5      & 280       & 120       & II        & II        & II        & II        & SN \\
		970508 & 20        & L         & 0.835     & 10.9      & 145       & 61.2      & II        & II        & II        & II        & --- \\
		970828 & 66.17     & L         & 0.9578    & 33.8      & 531       & 2620      & II        & II        & II        & II        & --- \\
		971214 & 15.9      & L         & 3.418     & 3.6       & 791       & 1460      & II        & II        & II        & II        & --- \\
		980425 & 18.05     & L         & 0.0085    & 17.9      & 55        & 0.01      & I         & II        & II        & II        & SN \\
		\hline                            
	\end{tabular}
	
	\tablefoot{Table \ref{table:rest} is published in its entirety in the machine-readable format. A portion is shown here for guidance regarding its form and content.\\
	\tablefoottext{a}{S and L represent Short GRBs and Long GRBs, respectively.}\\
	\tablefoottext{b}{I and II represent Type I GRBs and Type II GRBs, respectively.}\\
	\tablefoottext{c}{I and II represent GRBs-I and GRBs-II, respectively.}}
\end{table}
\end{appendix}

\end{document}